\documentclass{article}

\usepackage{arxiv}

\usepackage[utf8]{inputenc} 
\usepackage[T1]{fontenc}    
\usepackage{hyperref}       
\usepackage{url}            
\usepackage{booktabs}       
\usepackage{amsfonts}       
\usepackage{nicefrac}       
\usepackage{microtype}      
\usepackage{lipsum}
\usepackage{graphicx}
\usepackage{amsmath}
\graphicspath{ {./images/} }

\usepackage{amssymb}
\usepackage{amsthm}
\usepackage{mathrsfs}

\usepackage{algpseudocode}
\usepackage{soul}
\usepackage[inline]{enumitem}

\newcommand{\modulation}{\mathit{\Phi}_\mathrm{modulation}}
\newcommand{\aberration}{\mathit{\Phi}_\mathrm{aberration}}

\DeclareMathOperator*{\argmin}{arg\,min}

\begin{document}

\title{Differentiable optimization of the Debye-Wolf integral for light shaping and adaptive optics in two-photon microscopy}

\author{
 Ivan Vishniakou$^*$ \\
  Max Planck Institute\\ for Neurobiology of Behavior -- caesar\\ (MPINB) Bonn, Germany\\
   \And
 Johannes D. Seelig\thanks{Corresponding authors: ivan.vishniakou@mpinb.mpg.de, johannes.seelig@mpinb.mpg.de} \\
  Max Planck Institute\\ for Neurobiology of Behavior -- caesar\\ (MPINB) Bonn, Germany\\
}

\maketitle

\begin{abstract}
Control of light through a microscope objective with a high numerical aperture is a common requirement in applications such as optogenetics, adaptive optics, or laser processing. 
Light propagation, including polarization effects, can be described under these conditions using the Debye-Wolf diffraction integral.

Here, we take advantage of differentiable optimization and machine learning for efficiently optimizing the Debye-Wolf integral for such applications. 
For light shaping we show that this optimization approach is suitable for  engineering arbitrary three-dimensional point spread functions in a two-photon microscope.
For differentiable model-based adaptive optics (DAO), the developed method can find aberration corrections with  intrinsic image features, for example neurons labeled with genetically encoded calcium indicators, without requiring guide stars.
Using computational modeling we further discuss the range of spatial frequencies and magnitudes of aberrations which can be corrected with this approach.
\end{abstract}

\section{Introduction}

Control of light through high numerical aperture (N.A.) objectives is a common requirement in microscopy, for example for engineering specific point spread functions for super-resolution imaging \cite{bautista2016vector, liu2022super}, for generating target light distributions for optical stimulation \cite{lutz2008holographic, yang2011three, oron2012two}, for optical tweezers \cite{chen2013holographic, kumar2022manipulation}, or for aberration corrections in adaptive optics \cite{debarre2009image, ji2010adaptive, hu2020efficient}.
For controlling light in all these situations, computational modeling is the most versatile approach for finding a phase pattern that, when displayed on a spatial light modulator (SLM), results in the desired target light distribution in the focal volume. 

Light propagation through a microscope objective with high N.A. can accurately be described with the vectorial Debye-Wolf diffraction integral \cite{wolf1959electromagnetic}.  The Debye-Wolf integral takes into account the orientation of the electromagnectic field vector (polarization) which contributes to the shape of the focus in high N.A. objectives. Such effects can be exploited for high resolution imaging,  for example with diffraction limited objects, such as single molecules or nanostructures \cite{bautista2016vector, zhang2021single, liu2022super}.

However, inversion of the Debye-Wolf integral does not have a general closed-form solution \cite{foreman2008inversion}, and one therefore typically resorts to  numerical approaches for applications that aim to generate an intended target light distribution. 
A fast method for calculating the Debye-Wolf integral would therefore be useful across a range of applications, for example vectorial imaging \cite{torok2008high}, vectorial beam shaping for tight focusing \cite{liu2022super} or superresolution computational imaging \cite{foreman2011computational}, as well as any light shaping or imaging applications also at lower resolution, that is where polarization effects have less of an impact.

Here, we develop such an efficient approach for computing the Debye-Wolf integral using chirp-Z transforms (CZT) \cite{rabiner1969chirp} which can be used, both, for modelling (forward problem) and parameter search through optimization (inverse problem) within the Tensorflow  machine learning framework \cite{tensorflow2015-whitepaper}. This framework provides tools for differentiable optimization and leverages Graphics Processing Units (GPUs) for performance (see Sec. \ref{ssec:software}). Differentiable optimization of physical models is used across disciplines and is often called differentiable physics (see for example \cite{de2018end, schoenholz2020jax, thuerey2021pbdl}).

We demonstrate two applications of the implemented differentiable model: light shaping as well as adaptive optics in a two-photon fluorescence microscope, in  both cases by solving an inverse problem.

Model-based adaptive optics takes advantage of modeling image formation in the microscope for finding aberration corrections, for example by taking into account how aberrations affect the imaging system \cite{hanser2004phase, booth2006wave, debarre2009image, linhai2011wavefront, thao2020phase, vishniakou2020differentiable, vishniakou2021differentiable} or by introducing constraints on the structure of aberrations \cite{booth1998aberration, ji2010adaptive, song2010model}. Differential model-based adaptive optics (DAO), in addition to using a computational model that describes image formation, relies on differentiable optimization of the model in machine learning frameworks \cite{vishniakou2020differentiable, vishniakou2021differentiable}. 

Model-based AO approaches generally aim to reduce the number of measurements required for finding corrections compared with other sensorless iterative methods \cite{booth2006wave, booth2007wavefront, linhai2011wavefront, sherman2002adaptive, marsh2003practical, wright2005exploration, vellekoop2015feedback}. Reducing the number of required measurements improves the speed with which corrections can be found and reduces photobleaching or phototoxicity, both of which is of interest in particular for dynamic biological samples.

For using the Debye-Wolf diffraction integral in the differentiable optimization of a microscope model for adaptive optics \cite{vishniakou2020differentiable, vishniakou2021differentiable}, an aberration is included in the model as an unknown parameter. This parameter is then found using optimization, matching the model output to experimental measurements. 
We show that by additionally including the fluorescence distribution or object function of the sample as an optimization parameter, DAO can be used for arbitrary samples, without the requirement of guide stars, provided that sufficiently informative, or high-resolution, features are present. We demonstrate sensorless aberration detection under imaging conditions encountered in biological samples, such as in the fruit fly (\textit{Drosophila melanogaster}) or the zebrafish (\textit{Danio rerio}) expressing genetically encoded calcium indicators.
Using simulations we additionally discuss the limits in terms of magnitude and spatial frequency over which this optimization approach is expected to work.

In a second application we use the Debye-Wolf integral for light shaping in the focal volume, as commonly applied for example in the context of optogenetics \cite{oron2012two, emiliani2022optogenetics} or other photostimulation techniques \cite{yang2011three, lutz2008holographic}. These applications typically use computer generated holography based on an iterative approach such as Gerchberg-Saxton \cite{Gerchberg72}, which relies on forward and inverse beam propagation calculations usually implemented with Fourier transforms 
\cite{pozzi2018fast, yang2011three}. 
Here, applying optimization of the Debye-Wolf integral for light shaping, we generate  arbitrary light intensity distribution in the focal volume of a two-photon microscope.

\section{Methods} 
 
\begin{figure}[h!]
    \centering
    \includegraphics[width=1.0\textwidth,trim={0 0cm 0cm 0},clip]{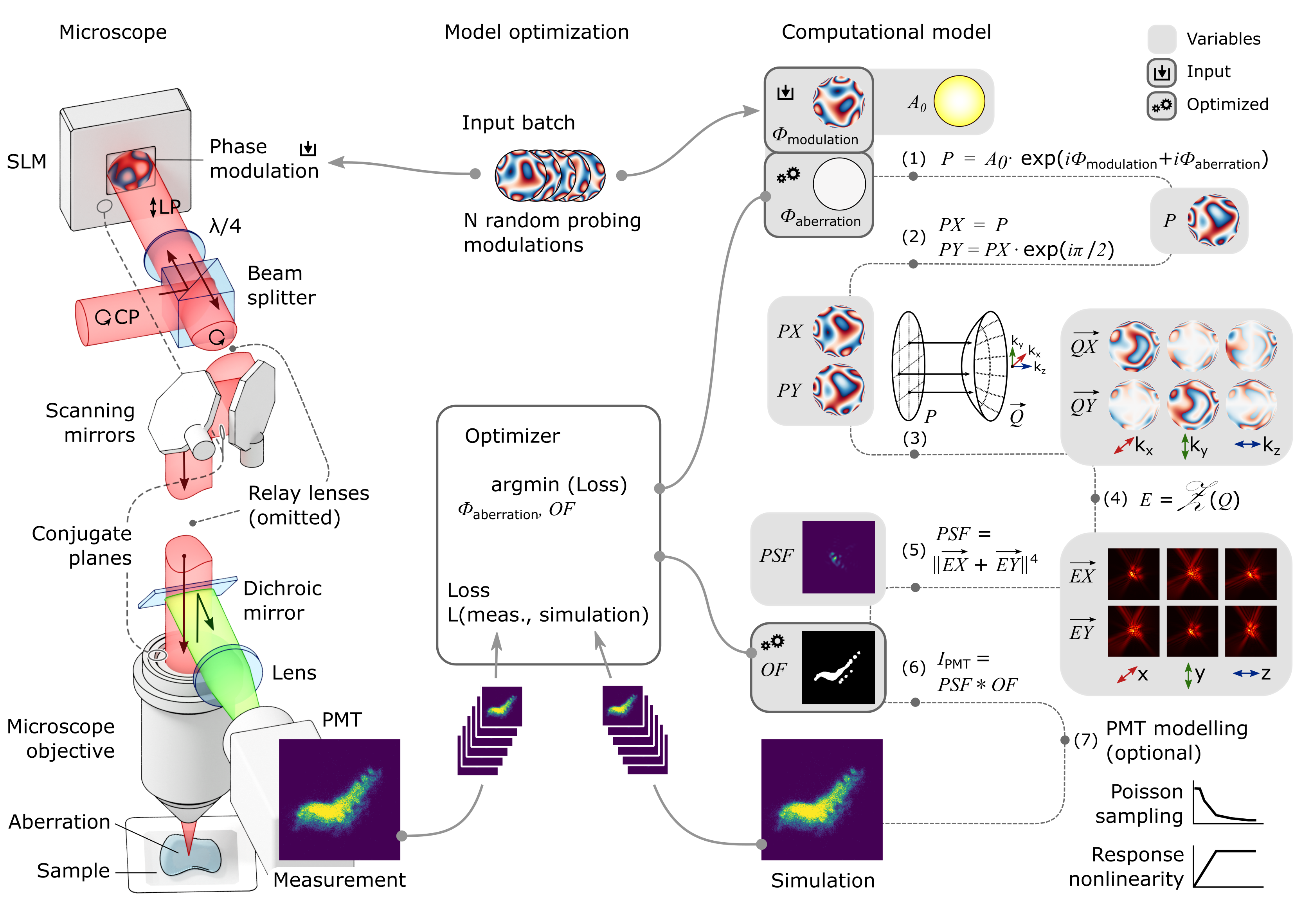}
    \caption{Schematic of microscope and DAO correction procedure. Left: schematic of the 2-photon microscope (relay lenses and mirrors omitted). An expanded circularly polarized beam is reflected off the phase-SLM (quarter waveplate ensures linear polarization of light aligned with SLM); scanned with mirrors and focused in the sample with an objective. Fluorescence is focused onto a PMT over a dichroic mirror. SLM, scanning mirrors and objective pupil are conjugate. Right: diagram of the computational model of the microscope. Taking SLM modulation as input, complex amplitude of the modulated beam at the SLM plane $P$ is calculated: (1) Incident polarized beam. (2) Pupil function is projected onto the focal sphere as in Debye-Wolf integral (3) resulting in angular spectra $Q$ for all three components of electromagnetic field  for either of x- and y- pupil polarization components. These are integrated with chirp-Z transform (4). The fluorescence PSF is obtained (5) and convolved with the object function $\mathit{OF}$ (6), which yields a simulated image. PMT and noise effects on image formation can also be included (7).\\
    Middle: diagram of model optimization to infer unknown aberration in the microscope. A number of random phase modulations is used to probe the microscope and corresponding fluorescence images are recorded. The model is run with the same modulations serving as input and optimized variables (object function and aberration) are tuned to match model output to the aberrated microscope by minimizing the loss function.
    }
    \label{fig:comp_model_scheme}
\end{figure}

The principle of the differentiable adaptive optics approach has been described in \cite{vishniakou2021differentiable} using the Rayleigh–Sommerfeld approximation \cite{shen2006fast}: a computational model of the microscope is created using wave optics, encompassing fixed parameters of the microscope such as N.A. and focal distance of the objective, controlled parameters, such as the applied illumination modulation, and unknown parameters, such as the aberration function. We here introduce and additional unknown parameter, the object function, which describes the fluorescence distribution, and makes the differentiable approach independent of guide stars (which were required in \cite{vishniakou2021differentiable}). This model functions as a digital twin of the microscope.
Once an unknown aberration is introduced in a sample in the microscope, the images produced by the model no longer match the experimental measurements. Given enough distinct measurements from the aberrated microscope with varying modulations (probes), the model output can be matched to the microscope by optimizing its aberration parameter (Fig. \ref{fig:comp_model_scheme}). Once the model output matches the probing images of the microscope, its aberration parameter can be transformed into a correction by taking the phase conjugate.

We here apply a similar approach but model image formation in the microscope by computing a vectorial Debye-Wolf diffraction integral of the pupil function at the focal plane, thus obtaining the point spread function of the microscope (PSF). The pupil function encompasses the apodization of the objective, applied SLM modulation, and system aberrations. The resulting PSF is convolved with the object function to obtain an image. 

\subsection{Setup}

We use a similar custom built two-photon fluorescence laser scanning microscope as in \cite{vishniakou2021differentiable}, equipped with a phase spatial light modulator (SLM) in the excitation path, as schematically shown in Fig. \ref{fig:comp_model_scheme}: an expanded laser beam (920 nm, Chameleon Discovery, Coherent), modulated by the SLM (Meadowlark, HSP1920-1064-HSP8) is focused onto the sample with the objective (Nikon CFI Apo 60X W NIR,  N.A. 1.0, 2.8 mm W.D.) and scanned with a pair of scanning mirrors (one resonant, one non-resonant). The SLM was imaged onto the resonant scanning mirrors with a pair of lenses and the mirrors were imaged into the back focal plane of the objective.
Fluorescence was collected with the same objective over a dichroic mirror collected with a photomultiplier (Hamamatsu, H7422PA-40) operating in photon counting mode. 
Images were recorded at a resolution of $512\times512$ pixels  at a rate of 30 Hz. The microscope was controlled with Scanimage \cite{pologruto2003scanimage} which was integrated with custom software written in Python for SLM control and for synchronizing probing modulations with image acquisition using Scanimage API callbacks.

\subsection{Computational model}
Vectorial diffraction theory (unlike scalar diffraction) accounts for the direction 
of the electromagnetic field and therefore allows to represent light polarization. This is relevant for modelling of high N.A. objectives, which deflect light by large angles and thus induce non-negligible oscillations of the electromagnetic field along the optical axis.
Modeling of image formation in the two-photon microscope 
proceeded along the following steps:

First, the pupil function of the objective was computed by representing it as a complex-valued function discretized at the chosen resolution. The amplitude of this function represents the beam profile, including apodization by the objective; the phase carries the modulation imposed by the SLM as well as the aberration of the sample, since aberrations are modeled at the plane conjugate to the SLM \cite{vishniakou2021differentiable}. This alleviates the necessity to model intermediate planes in the light propagation path and allows to convert aberrations determined through optimization (see below) into corrections by simple inversion of the phase $\varphi_{\mathrm{corr}}=-\varphi_{\mathrm{aberration}}$.
    
Second, simulation of the point spread function (PSF) corresponding to the pupil function was done by computing the diffraction of the wavefront at the focal plane considering objective and microscope parameters. 
The PSF was computed at the same scale $\mathrm{um}/\mathrm{px}$ as the microscope image discretization to ensure pixelwise matching of the simulated images. The PSF for two-photon imaging was obtained  by squaring the intensity of the excitation light. 
    
Next, image formation in the microscope was modelled as the convolution of the PSF and an object function. As for the PSF, the resolution of the object function matched the imaging resolution of the microscope.
    
Modelling gain and nonlinearity of the photomultiplier operating in photon counting mode is relevant to to get realistic images. The limited dynamic range of the PMT, which leads to saturation at high photon count rates, 
affects especially images of brighter objects, which as a result appear bigger. This process is modelled by multiplying the image intensity by the gain factor and capping the result at a constant value, corresponding to the maximum photon count rate. 
    
Two types of PMT noise were taken into account: shot noise (due to probabilistic nature of single photon counting) and dark current (thermally generated electrons in the PMT). Shot noise was modelled by drawing Poisson-distributed random variables with $\lambda$ proportional to the imaged intensity. Dark current was modelled by adding a positive bias to the image before the Poisson sampling. 

Combining all of the above steps, the model produced images that closely matched experimental data as shown in Fig. \ref{fig:comp_model_scheme}. More details of the implementation are provided in the next section.

\subsection{Efficient implementation of the Debye-Wolf integral optimization}

Using vectorial diffraction theory as described in \cite{boruah2009focal, leutenegger2006fast} requires the independent calculation of diffraction of the 3 orthogonal components of electromagnetic field.
We follow the convention in \cite{boruah2009focal, leutenegger2006fast} where the optical axis of the system is directed in z-direction of a Cartesian coordinate system. The electromagnetic field oscillations in a collimated beam are occurring in the x- and y- directions. The pupil function is then defined by two complex-valued distributions $\mathit{PX}$ and $\mathit{PY}$, corresponding to the x-, and y-polarized light field components. In our case, the beam is circularly polarized and therefore the complex amplitudes of both polarization components of the beam are identical except for a global $\pi/2$ phase difference. If $A_0$ is the amplitude profile of the beam, $\aberration$ is the aberration equivalent at the SLM plane, and $\modulation$ is the phase modulation imposed by SLM,
\begin{equation*}
\begin{split}
\mathit{EX} & = A_0\cdot\exp{(i\aberration+i\modulation)} \\
\mathit{EY} & = \mathit{EX}\cdot\exp{(i\pi/2)}.
\end{split}
\end{equation*}
Next, the PSF of the obtained pupil function is computed. This can be done by mapping the pupil function onto the converging spherical wavefront that the objective creates, which results in the Debye-Wolf integral \cite{wolf1959electromagnetic}, which has to be computed for every point of the target (focal) plane to obtain the PSF. This has to be solved numerically, and, depending on the resolution of the integrand and the target plane, naive calculation of the this integral is prohibitively slow. Additionally, the discretization resolution of the integrand needs to be high for strong phase perturbations to satisfy the sampling criterion (the phase difference between adjacent sampling points of the pupil function cannot exceed $\pi$ radians \cite{leutenegger2006fast}). However, this integral can be formulated in terms of Fourier transforms, if the integrand is suitably parameterized and sampled, allowing to leverage the fast Fourier transform (FFT) algorithm with lower time complexity, which is the cornerstone of the angular spectrum method \cite{shen2006fast}.
For that, the entrance pupil function is mapped into the angular spectrum -- a distribution of complex amplitudes over k-space, where each point represents a plane wave with given amplitude and phase travelling in a distinct direction \cite[Chapter~3]{goodman2005introduction}. In this representation, the focal plane diffraction is a Fourier transform of the angular spectrum. Similarly, diffraction in other offset planes can be computed if the angular spectrum is multiplied by a corresponding propagation factor, which allows simulating focal volumes.

For calculating the angular spectrum, each of the x- and y-polarizations of the objective pupil functions is remapped onto 3-dimensional k-space $Q(k_X, k_Y, k_Z)$, sampled over the region corresponding to the angle of convergence of the focused beam. Diffraction of this 3D pupil function at the target plane is expressed with a Fourier transform (taken component-wise), denoted here as $\mathcal{F}$: 
\begin{equation*}
\begin{split}
\mathit{EX}_{X,Y,Z} &= \mathscr{F}[QX_{k_{X},k_{Y},k_{Z}}] \\
\mathit{EY}_{X,Y,Z} &= \mathscr{F}[QY_{k_{X},k_{Y},k_{Z}}].
\end{split}
\end{equation*}
The resulting vector electric field in the target plane $E=(E_X, E_Y, E_Z)$ is found by summing the corresponding components of the vectorial diffraction of either of X- and Y- polarization components of the beam:
\begin{equation*}
\begin{split}
E_X &= EX_X + EY_X \\
E_Y &= EX_Y + EY_Y \\
E_Z &= EX_Z + EY_Z.
\end{split}
\end{equation*}

We use the mapping of $\mathit{EX}$ and $\mathit{EY}$ into k-space $\mathit{QX}$ and $\mathit{QY}$ from \cite{boruah2009focal}. However, using FFT for integrating the angular spectrum has the limitation that the resolution of the output is the same as the resolution of the input. As a result, the frequency band of the integrand results in a major part of the k-space remaining empty, although it still needs to be computationally processed, while the relevant frequency band has only very low resolution \cite{hu2020efficient} (see Supplementary Materials).
Instead of FFT we therefore use the chirp-z transform (further denoted as $\mathscr{Z}$), which is free of these constraints, allowing to arbitrarily chose the resolution of the output and the frequency band of the input, alleviating the idle computations and keeping the resolution of the input pupil function high \cite{leutenegger2006fast}. Using Bluestein's algorithm, it can be computed with two FFTs and one multiplication with a precomputed coefficient matrix \cite{rabiner1969chirp}.
This allows to freely select the resolution of the discretization of the pupil function, as well as the size of the calculated field of view and its resolution. This in turn allows to easily adjust the model to parameters of the microscope, such as zoom level or imaging resolution \cite{hu2020efficient}.

The measured light intensity is proportional to the squared magnitude of the electric field (considering all 3 directional components), and taking into account the two-photon fluorescence excitation process, we obtain the fluorescence PSF as:
\begin{equation*}
\begin{split}
\mathit{PSF} &= (|E_X|^2 + |E_Y|^2 + |E_Z|^2)^2\\
\end{split}
\end{equation*}
Given an object function $\mathit{OF}$, the fluorescence image $I_\mathrm{fluorescence}$ is obtained as the convolution of the object function with the PSF: $I_\mathrm{fluorescence}=\mathit{OF}*\mathit{PSF}$. This assumes that the entire field of view lies within the isoplanatic patch.

Finally, to simulate photon counting, the measured intensity in photon counts for every pixel is sampled from a Poisson distribution $\mathrm{Pois(\lambda)}$ with $\lambda \propto I_\mathrm{fluorescence}$. To control the proportionality we introduce the gain parameter $a$, simulate background noise with a global positive bias to the fluorescence $b$ before Poisson sampling, and the maximal photon count is capped at $I_\mathrm{max}$:
$$I_\mathrm{PMT} = \min({\mathrm{Pois}(a \cdot I_\mathrm{fluorescence} + b), I_\mathrm{max}})$$
This allows to closely match actually recorded focal volumes of the microscope once all the parameters are fitted.
We verified that the implementation of the Debye-Wolf integral accurately reproduced vectorial PSFs as described in \cite{leutenegger2006fast}, see Supplementary Materials.

\subsection{Fitting the model to the experimental setup}

Parameters for fitting the model to the microscope included N.A. (refractive index of immersion medium, pupil diameter and focal length of the objective), illumination wavelength,  as well as size and resolution of the field of view. These parameters need to be adjusted such that the resulting PSF generated computationally  dependent on an arbitrary phase image  matches the experimentally measured PSF when displaying the same phase pattern on the SLM.

The procedure for model matching was similar to the one previously described in \cite{vishniakou2021differentiable}. We started with the physical parameters of the microscope and fine-tuned them manually to match the model response to low-order Zernike modes to that of the microscope when imaging $0.5\mu \textrm{m}$ fluorescent beads. 
Compared with the physical parameters, only a minor adjustment of the numerical aperture was needed, achieved by reducing the pupil diameter, since the microscope objective was slightly underfilled. Displaying phase ramps of varying magnitudes (tip and tilt modes Z2 and Z3) caused focus displacements in the field of view depending on the numerical aperture (see Supplementary Materials Fig. \ref{itm:model_matching_illustration}). By adjusting the model numerical aperture, the lateral displacements of the foci were matched to the measured ones (see Supplementary Materials).

Additionally, the correspondence between the AO system of the microscope and the model needed to be adjusted. The displayed patterns were slightly rotated to match the geometry of the setup 
which was achieved by adding the rotation to the phase patterns before displaying them on the SLM. Furthermore, the modulations applied to the SLM needed to exactly match the beam diameter and any light entering the system not accounted for in the model had to be blocked. For this, we implemented binary amplitude modulation with the phase-only SLM \cite{mendoza2014encoding} and muted the SLM area outside the modulation, effectively creating an aperture with the SLM and allowing only modulated light to proceed to the sample.
The beam intensity distribution within the modulated area was obtained through optimization of the model in the same way as as used for finding aberrations, and embedded in the model as amplitude of the pupil function (apodization).

The same amplitude modulation technique was also used to match the simulated PMT gain parameters: a number of phase modulations with varying attenuation from 1.0 to 0.0 were consecutively tested in the microscope. The gain was set such that overexposure (and PMT photon count capping) happened at full amplitude; this was indicated by the focal spot getting broader without its maximum count number increasing. At a particular level of attenuation, the imaged focus reached its maximum before being capped, and the model parameters (gain level and maximum photon count) were adjusted to reach the cap at the same level of beam attenuation.

\subsection{Model optimization}

Once the model is initialized with correctly adjusted parameters, it can be treated as a function generating fluorescence images depending on multiple variables, such as SLM modulation, sample aberration, and object function: $I_\mathrm{PMT} = M(\modulation, \aberration, \mathit{OF})$. Finding the unknown parameters then can be formulated as an optimization problem of matching the output of the model to the observations:
$$
\argmin_{\mathit{OF}, \aberration} \sum_{i=1}^{N}\mathrm{loss}(M(\mathit{\Phi}_{\mathrm{modulation}\ _i}, \aberration, \mathit{OF}), I_{\mathrm{i}}),
$$
where $I_i$ is a fluorescence image taken with the microscope under the i-th out of $N$ probing modulations, $\modulation\  _i$, and the loss is the function measuring the difference between images, which is to be minimized.
The loss function was constructed from the following terms:
$$
\mathrm{loss} = -r(M(\modulation, \aberration, \mathit{OF}), I_\mathrm{measured}) - I_\mathrm{max} + L_1(\mathit{OF}),
$$
where $r$ is Pearson's correlation coefficient between simulated images and measurements. $I_\mathrm{max}$ is the maximum intensity of the PSF which promotes solutions that do not discard light out of the field of view. The $L_1(\mathit{OF})$ norm promotes sparsity of the object function with the goal of getting a minimal footprint of objects and thus imposing the assumption that image blurring (and visible object footprint expansion) happens due to PSF broadening.

For object function optimization, an initial estimate was selected based one the probing images by taking the brightest of all measurements.
(In situations where the object is known and can be accurately modelled, for example when guide stars of known size are present in the sample, this information could be used alternatively.)
 
The model was implemented in Tensorflow with the following adjustments:
First, the PMT shot noise simulation was omitted, since the random sampling operation interrupts gradient backpropagation needed for optimization. Also, pixel-wise image comparison in the loss function becomes very irregular for noisy low-intensity images. Second, the PMT nonlinearity simulated with a $\mathrm{max}$ function was implemented with a repurposed leaky ReLU function to prevent gradient loss in high-intensity areas and to facilitate the optimization process.
Third, the object function $\mathit{OF}$ and PSF convolution was implemented as a multiplication in the Fourier domain, since direct convolution in Tensorflow is very resource exhaustive for big kernels as used here. Fourth, the chirp-Z transforms were optimized and sped up by precomputing and storing the coefficient matrices and sharing them across multiple transforms, since these coefficients depend on the respective resolution of the transform input and output only, and not on the transformed data.
Fifth, following model evaluation, the maximal intensity $I_\mathrm{max}$ of the PSF was saved to be used in the loss function. Finally, simulated and recorded images were compared over a cropped central region. The reason was that SLM modulations can cause images to shift substantially in the field of view, causing ambiguity at the edges. Cropping images to the central region means that the object function was simulated for a greater area than was typically visible, allowing to reproduce images shifting without edge artifacts.

\subsection{Model optimization for adaptive optics}

For adaptive optics the goal is the identification of an unknown aberration as well as an unknown object function (unless guide stars are present in the sample). To constrain the optimization problem, $N=56$ random phase aberrations were generated by randomly sampling the first $55$ Zernike coefficients from a normal distribution and by taking the weighted sum of the corresponding polynomials. The magnitude of the resulting probe phase modulation was normalized to $\mathit{RMS}=0.18\lambda$. Applying these modulations consecutively, fluorescence images were recorded under aberrated conditions, resulting in a dataset of pairs of input modulations (probes) and corresponding output images (responses).
Out of all these response images, the brightest one was selected and assigned to the object function variable as an initial estimate; the inverse of the corresponding probing modulation was assigned as an initial estimate for the aberration.

Optimization was done with Adam with default parameters and a learning rate of $0.1$. We did not use the standard training loop which iterates through a complete dataset introducing each sample once per epoch. Instead we implemented a custom loop which draws random samples from the dataset to populate a batch in every iteration and therefore does not operate in terms of epochs:
\begin{algorithmic}
\State $\mathrm{i\_init}=\mathrm{indexof(max(brightness(images)))}$
\State $\mathrm{aberration} \gets \mathrm{-modulations[i\_init]}$
\State $\mathrm{object\_function} \gets \mathrm{images[i\_init]}$
\For{$(i=0$; $i<\mathrm{max\_iter}$; $i)$} 
    \State $\mathrm{modulations}, \mathrm{images} \gets \mathrm{sample\_random\_batch(modulations, images)}$
    \State $\mathrm{simulated} \gets M\mathrm{(modulations, aberration, object\_function)}$
    \State $\mathrm{gradients} \gets \mathrm{get\_gradients(loss(simulated, images))}$
    \State $\mathrm{aberration, modulation} \gets \mathrm{apply\_gradients(aberration, modulation, gradients)}$
\EndFor
\State $\mathrm{correction} \gets -\mathrm{aberration}$
\end{algorithmic}
Typically, we used $300$ such iterations with a batch size of $8$ and the solution was obtained once the correlation between probes and simulations exceeded $0.5$. 
Additionally we did not have validation data since we were dealing with a very small number of samples (<100). 
  
\subsection{Model optimization for light shaping}

Since the model is optimized for matching an output to a target, it can also be similarly used for light shaping. In this case, a synthetic image serves as target. Since the shape of the focus  has to match the synthetic target image, the object function is not contributing to image formation and can be set to a Dirac delta function or a small bead, without being subject to optimization. Additionally, since no probing modulations are required, the optimization is done with batch size equalling one (flat phase mask).

\subsection{Simulation of PSFs depending on different aberration conditions}

Aberrations encountered in biological samples result from sample inhomogeneities, or changes in refractive index, across spatial scales, from variations between individual cells to changes in tissue composition. Such inhomogeneities lead to changes in the optical path  between different parts of the beam propagating in the sample and thus impede focusing of light to a diffraction limited spot.
To assess the range of aberrations that one can expect to correct with the developed approach, we simulated and quantified the effect of various aberrations with varying spatial frequencies and phase magnitude on the PSF and consecutively imaging quality. 

We use conventions from aberration theory \cite{gross2007handbook} and describe an aberration as optical path difference (OPD) of an aberrated compared to an unaberrated system, measured in wavelengths $\lambda$. Here, this corresponds to the deviation of the wavefront from an ideal spherical shape, comprising the OPD surface $W$. This surface can be characterized by its magnitude, for example peak-to-valley difference $W_\mathrm{PV}$ between the maximal (peak) and minimal (valley) OPD of the surface, or $W_\mathrm{RMS}$ - root-mean-square of the OPD, and by its spatial frequency $\nu$ - number of peak-valley cycles across the pupil, which roughly translates to the size of phase discontinuities in the aberrating medium.
Either parameter has an effect on the resulting PSF, reducing it's peak intensity by reducing constructive interference through off-phasing parts of the wavefront, or introducing destructive interference with bigger magnitudes, which splits the focus and adds speckles. The effect of the latter and the complexity of the speckle patterns depend on the spatial frequencies of the aberration. 

We assessed both effects by simulating the effect of randomly drawn aberrations with $W_\textrm{RMS}$ and $\nu$ on the PSF in our microscope model by measuring the resulting Strehl ratio $S$, resulting in a map of aberrations allowing to qualitatively describe their effect on the focus (Fig. \ref{fig:aberration_map}). The aberrations were generated by integrating a single frequency-band angular spectrum with randomized phase using CZT, and normalizing the surface magnitude to $W_\mathrm{RMS}$.

\subsection{Sample preparation}
Flies expressed the genetically encoded calcium indicator jGCaMP8m \cite{zhang2021fast} in ensheathing glia using the GAL4 line R56F03 \cite{jenett2012gal4} and were dissected using laser surgery and glued to a glass cover slide as described in \cite{flores2022dynamics}. 
Zebrafish (\textit{Danio rerio}) Tg(elavl3:GCaMP6s) 5dpf larvae expressed GCaMP6s pan-neuronally \cite{ahrens2012brain}, and were fully embedded in 3.5\% low-melting agarose.

\section{Results}

\subsection{Differentiable adaptive optics using the Debye-Wolf integral}

To validate the performance of the developed DAO approach  we used several different samples: fluorescent beads (with a diameter of 0.5 $\mu$m) distributed in clusters on a cover glass, as well as fruit flies (\textit{Drosophila melanogaster}) and zebrafish larvae (\textit{Danio rerio}). None of these samples included guide stars and optimization instead took advantage of sample intrinsic features. Fruit fly and zebrafish samples were largely transparent, without inducing any visible distortions of the PSF shape. 
Aberrations were therefore introduced by applying a phase pattern generated from the first 56 Zernike modes on the SLM. 
The samples were then imaged under these aberrating conditions and probed with 56 randomly selected modulations with the same maximal order of Zernike modes. 
This procedure had the advantage of that the unaberrated ground truth was known and therefore could be compared with the achieved correction. 

We additionally verified that the developed approach could correct for aberrations generated by aberrating samples, such as a layer of vacuum grease on top of fluorescent beads (in which case the ground truth was not known). 
The intrinsic aberrations encountered in zebrafish and fruit flies, which lead  to reduced intensity when imaging deeper into the sample, exceeded the range of applications of the method, likely consisting of very high frequency aberrations that lead to dimming of the excitation light, without however leading to detectable changes in the PSF shape (see section \ref{ssec:abb_characterization}).

\begin{figure}[h!]
    \centering
    \includegraphics[width=1.0\textwidth, trim={0 0cm 0cm 0},clip]{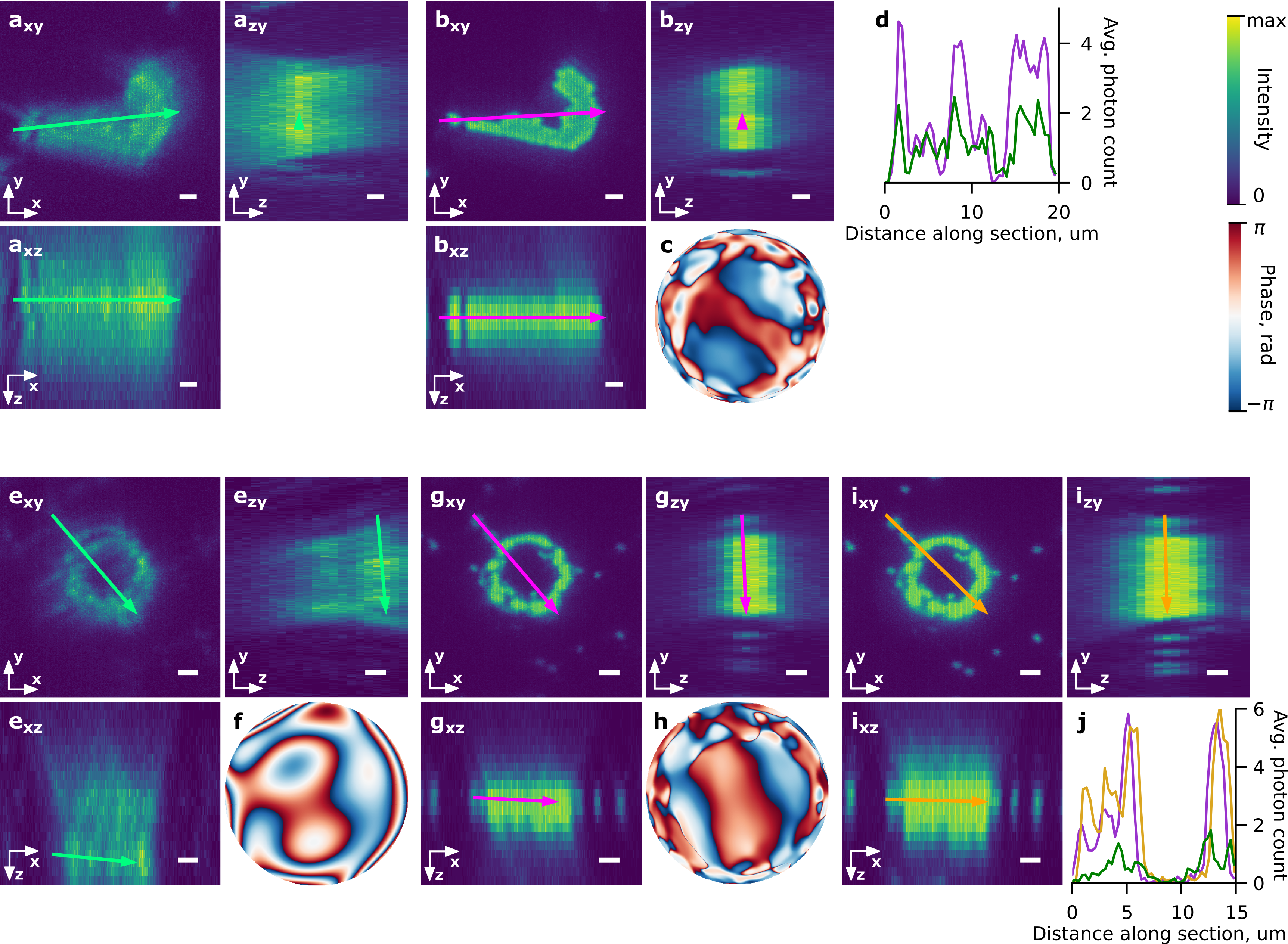}
    \caption{Aberration correction with sample of aggregates of 0.5 $\mu$m diameter fluorescent beads on a cover glass. Top half: sample aberrated by a layer of vacuum grease. Panels a$_\mathrm{xy}$, a$_\mathrm{zy}$, a$_\mathrm{xz}$ - max projections of the sample volume imaged without correction, b$_\mathrm{xy}$, b$_\mathrm{zy}$, b$_\mathrm{xz}$ - with correction. c - found aberration, and d - intensity cross sections (color coded) from uncorrected and corrected stacks. \\
    Bottom half: Aggregate of 0.5 $\mu$m fluorescent beads on cover glass, aberration is imposed by the SLM. Panels e$_\mathrm{xy}$, e$_\mathrm{zy}$, e$_\mathrm{xz}$ - max projections of the sample volume taken with aberration; g$_\mathrm{xy}$, g$_\mathrm{zy}$, g$_\mathrm{xz}$ - with correction; i$_\mathrm{xy}$, i$_\mathrm{zy}$, i$_\mathrm{xz}$ - without aberration. f - imposed aberration, h - sensed aberration, j - plots of intensity (color coded) of the shown sections. All scale bars are 2 $\mu$m. 
     }
    \label{fig:reults_1}
\end{figure}

\begin{figure}[h!]
    \centering
    \includegraphics[width=1.0\textwidth, trim={0 0cm 0cm 0},clip]{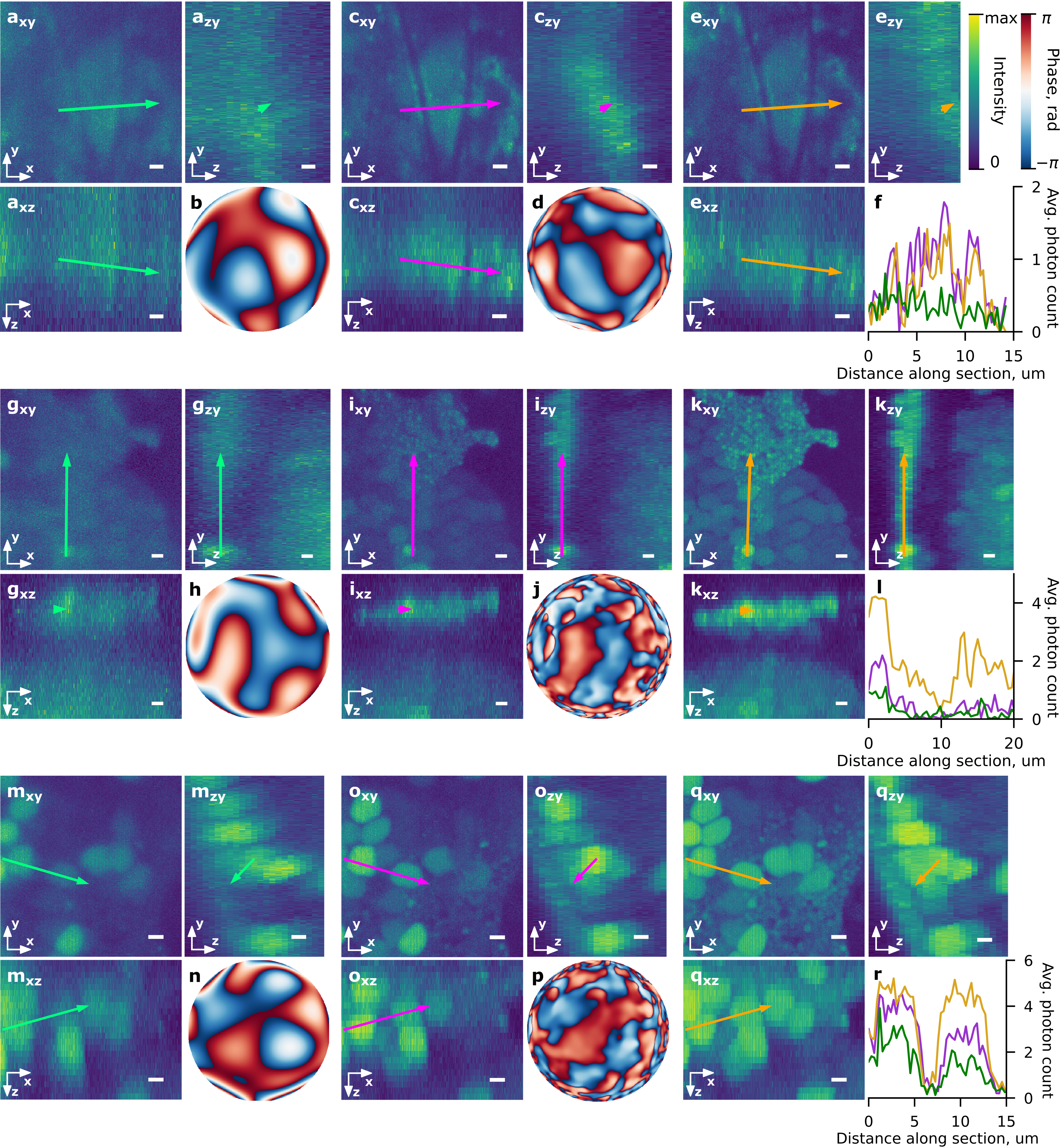}
    \caption{Correction of SLM-imposed aberrations in biological samples. Top group: fruit fly \textit{Drosophila melanogaster}. a$_\mathrm{xy}$, a$_\mathrm{zy}$, a$_\mathrm{xz}$ - max projections of the sample volume imaged without correction; c$_\mathrm{xy}$, c$_\mathrm{zy}$, c$_\mathrm{xz}$ - with correction; e$_\mathrm{xy}$, e$_\mathrm{zy}$, e$_\mathrm{xz}$ - without aberration. b - imposed aberration, d - found aberration, and f - intensity cross sections (color coded) from uncorrected, corrected and unaberrated stacks. \\
    Middle group: zebrafish sample, g$_\mathrm{xy}$, g$_\mathrm{zy}$, g$_\mathrm{xz}$ - max projections of the sample volume imaged without correction; i$_\mathrm{xy}$, i$_\mathrm{zy}$, i$_\mathrm{xz}$ - with correction; k$_\mathrm{xy}$, k$_\mathrm{zy}$, k$_\mathrm{xz}$ - without aberration. h - imposed aberration, j - found aberration, and l - intensity sections (color coded) from uncorrected, corrected and unaberrated stacks.\\
    Bottom group: another zebrafish sample, m$_\mathrm{xy}$, m$_\mathrm{zy}$, m$_\mathrm{xz}$ - max projections of the sample volume imaged without correction; o$_\mathrm{xy}$, o$_\mathrm{zy}$, o$_\mathrm{xz}$ - with correction; q$_\mathrm{xy}$, q$_\mathrm{zy}$, q$_\mathrm{xz}$ - without aberration. n - imposed aberration, p - found aberration, and r - intensity sections (color coded) from uncorrected, corrected and unaberrated stacks.
    }

    \label{fig:results_2}
\end{figure}

As seen in examples with fluorescent bead clusters (Fig. \ref{fig:reults_1}), fruit flies and zebrafish (Fig. \ref{fig:results_2}) corrections led to an increase in image contrast and intensity.
However the corrected PSF brightness was generally below the unaberrated ground truth. This was likely due to high spatial frequency noise,  resulting from measurement noise, manifesting itself as high-order aberrations similar to what is discussed in \cite{hu2020universal}. Model optimization required typically around 120 seconds on a workstation with four Nvidia Titan RTX GPUs.

\begin{figure}[h!]
    \centering
    \includegraphics[width=1.0\textwidth,trim={0 0cm 0cm 0},clip]{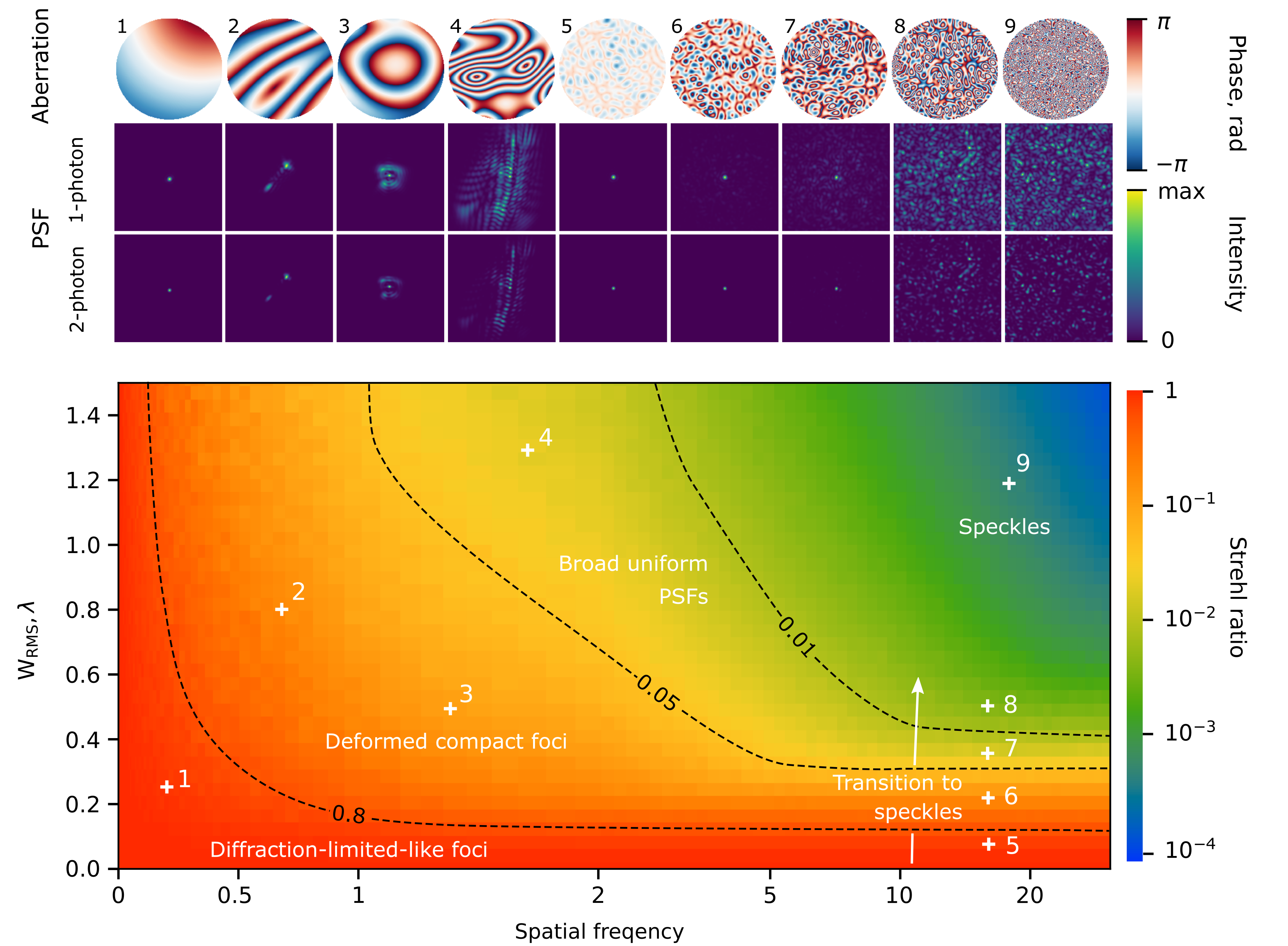}
    \caption{Simulation of aberrations with varying magnitude $W_\textrm{RMS}$ and spatial frequency, averaged over 50 repetitions with representative aberrations and corresponding single- and two-photon PSFs for each point of the parameter space. Simulated field-of-view width equals $20\mu\mathrm{m}$. Isolines separate the characteristic regions of the parameter space. }
    \label{fig:aberration_map}
\end{figure}

\subsection{Aberration characterization} \label{ssec:abb_characterization}

To investigate the domain over which aberrations can be corrected with the developed model, 
we characterized the effect of aberrations on the microscope PSF. For this, we quantified aberrations by their spatial frequency and phase magnitude and explored a range of aberrations by running a simulation of 50 random aberrations per data point and averaging the resulting PSF Strehl ratio $S$. 

The resulting map of the parameter space confirms that both high phase magnitude and high spatial frequency correlate negatively with focus brightness. One can qualitatively describe several regions of the parameter space (Fig. \ref{fig:aberration_map}):\
First, bright foci with $S>0.8$ (samples 1, 5) which appear as close to diffraction-limited can be achieved even at high magnitude aberrations with low enough spatial frequency (where aberrations are ramp-like, causing only the displacement of the focus), or at high spatial frequencies when 
 aberration magnitude is low. In a second region with compact deformed foci (samples 2, 3) corresponding to lower-order Zernike modes - often the  shapes of aberrations can be recognized (coma, trefoil, astigmatism, etc.).
Third, a region with broader foci expanding over or beyond the whole field of view, at higher spatial frequencies or magnitudes (sample 4).
Fourth,  a speckle regime (samples 8, 9) with no recognizable focus shapes, only uniformly scattered speckles in and beyond the field of view.
Finally, a transition to a region dominated by speckles at above 10 cycles across the pupil. Further increasing the spatial frequency did not affect the quality of the PSF. The PSF depends only on the magnitude of the aberration and shows a characteristic transition with increasing magnitude (samples 5 through 8): the focus becomes dimmer (5, 6), then  coexists with appearing background speckle (7) and gets even dimmer while the background speckle gets brighter, until it can not be distinguished from the speckles (8).

These qualitative observations describe several imaging situations: low spatial frequency aberrations, such as spherical, coma, or astigmatism, resulting for example from a refractive index mismatch at the sample interface, can have a strong deteriorating effect on the image quality by deforming the PSF. They can however be relatively easily corrected, 
since the number of degrees of freedom for the correction is low.
Regions in Fig. \ref{fig:aberration_map} with high spatial frequencies and a transition of the focus to speckles are reminiscent of imaging deep in tissue: with depth, image brightness reduces without affecting the resolution at first. As background speckles increase in brightness, the image contrast is lost until no details are distinguishable (since the speckled PSF may become as big as the FOV). The advantage of the nonlinearity of multi-photon imaging is especially evident here, since it increases the contrast of a dim focus and the background speckles for higher magnitude aberrations.

The aberrations in these simulations were generated with a single spatial frequency band, whereas in actual samples aberrations with multiple frequencies can overlap, presenting additional challenge and possibilities of partial corrections, for example when a low-frequency component is corrected and sensed, while a high-frequency residual aberration stays, since it can not be measured.

The aberrations used in our experiment were falling in the range of up to 2 cycles across the phase image, and these strongly affect the focus shape, and therefore could be sensed by our image-based method. 
From the simulation results we expect our method to work for aberrations with spatial frequencies up to around 5 cycles across and with magnitudes limited by the focus brightness and measurements signal-to-noise ratio or the size of the focus. Increasingly more probes are required for higher spatial frequency aberrations. This matches the conclusions from \cite{hu2020universal}: more complex aberrations require probing with higher spatial frequency component and also are more susceptible to noise.

The speckle range of the parameter space is the most challenging to correct, since apart from the large number of degrees of freedom, these aberrations allow only a very small isoplanatic patch. 
These conditions require significantly more (intensity-only) measurements, and can be addressed by methods such as \cite{tang2012superpenetration, papadopoulos2017scattering, may2021fast}.

\subsection{Light shaping}

\begin{figure}[h!]
    \centering
    \includegraphics[width=1.0\textwidth, trim={0 0cm 0cm 0},clip]{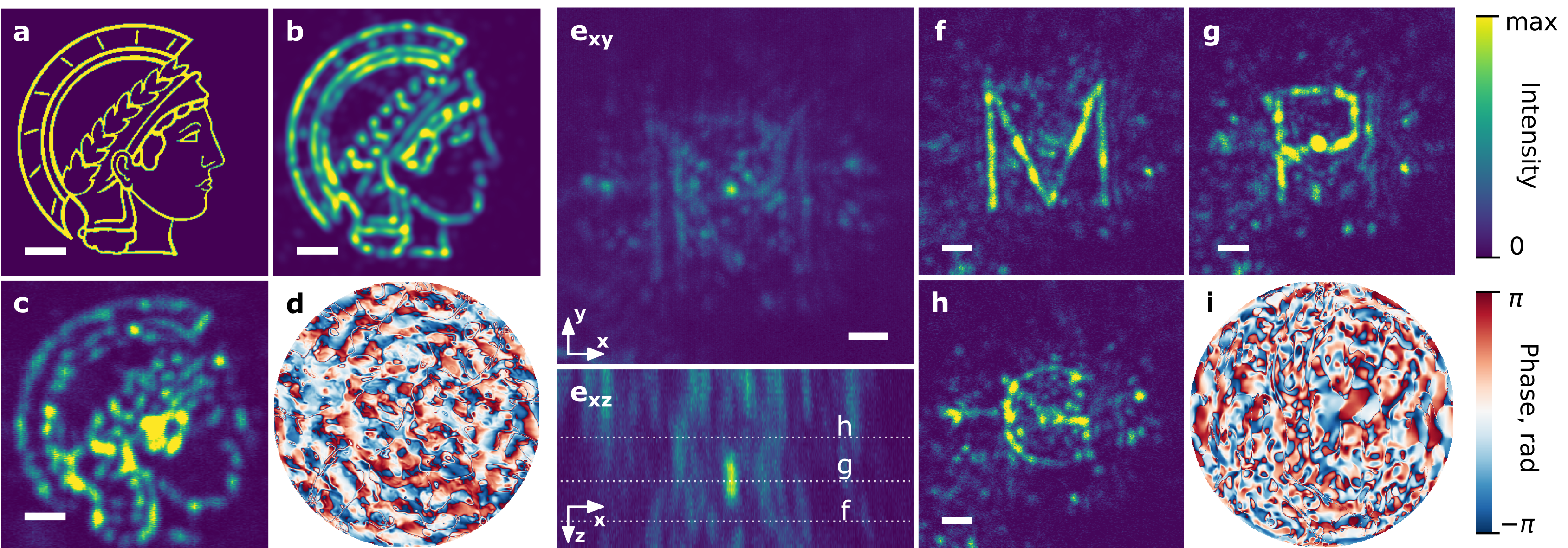}
    \caption{PSF engineering with differentiable microscope model. Left group: planar PSF with image of Minerva in a $20 \times 20 \mu$m FOV, where . (a) Desired PSF intensity distribution (a synthetic raster image). (b) Result of model PSF optimization. (c) Microscope image of a $0.5 \mu$m fluorescent bead with the engineered PSF. (d) Obtained phase modulation for the desired PSF. Right group: volumetric PSF in shape of letters `MPG` at different depths; (e$_\mathrm{xy}$) and (e$_\mathrm{xz}$) are max projections of the focal volume on xy and xz axes respectively. (f), (g), and (h) are xy- sections taken at different depths, as marked in (e$_\mathrm{xz}$). (i) Beam phase modulation for the PSF. All scale bars are $2\mu$m.}
    \label{fig:result_hologram}
\end{figure}

We demonstrate light shaping using differentiable model optimization by generating planar (in a single focal plane) and volumetric holograms (in three focal planes at the same time) and by testing them in the microscope. For this, we used single beads of 0.5 $\mu$m diameter, which allows to image the PSF. Note that since such PSF shaping requires redistribution of energy of the beam over a larger volume, the intensity of the two-photon excitation dramatically reduces. A fraction of the light that reflects off the SLM cover glass was not modulated and becomes relatively strongly visible (in the center of Fig. \ref{fig:result_hologram}e). 

\subsection{Software}
\label{ssec:software}

Python code and detailed instructions with examples are provided in  Supplementary Materials. All the underlying functions for the model are available in Tensorflow, thus allowing to implement the simulation with the provided automatic differentiation, gradient-based optimizers and GPU-acceleration. The model can be used either for simulation of the microscope and generation of synthetic images, or for wavefront sensing by fitting it to the output of the aberrated microscope.

All code used for the applications described in the current paper will be provided upon publication at\\
\texttt{https://github.com/ivan-vishniakou/differentiable-adaptive-optics}, and a Jupyter notebook with examples is provided in Supplementary Materials. The code is split into three dedicated Python modules: \texttt{czt\_tf.py}, containing the implementation of the chirp-z transform in Tensorflow, \texttt{microscope\_model.py} with the end-to-end microscope model, and \texttt{wavefronts.py} with helper functions to generate wavefront shapes from Zernike polynomials. Additionally, Jupyter Notebooks are provided with examples of how to run the model optimization and for synthetic image generation (\texttt{microscope\_demo.ipynb}), vectorial PSF validation (\texttt{vectorial\_psf.ipynb}), as well as demonstration of the chirp-Z transform (\texttt{czt\_demo.ipynb}).

\section{Discussion}

We developed a framework for differentiable modeling of light propagation with the Debye-Wolf diffraction integral. Efficient computation and optimization as well as close integration with an experimental setup opens up a range of applications. While we used the developed framework for DAO as well as light shaping, the same method could similarly be used in situations where the vectorial aspects of the computation are more important \cite{foreman2011computational, liu2022super}. 

Different from our previous DAO implementation \cite{vishniakou2020differentiable, vishniakou2021differentiable}, which used Rayleigh–Sommerfeld diffraction \cite{shen2006fast}, we here used the Debye-Wolf integral, which describes light propagation through high N.A. microscope objectives. Apart from being physically more accurate, an additional advantage is that the physical parameters of the objective can be directly inserted into the model and only minor parameter adjustments are needed for accurately describing image formation. Adjusting the model to the experimental setup is therefore simpler than when using the Rayleigh-Sommerfeld approximation \cite{vishniakou2021differentiable}.

Including an object function in the optimization allowed using intrinsic image features for DAO, that is, without requiring guide stars as in our previous implementation \cite{vishniakou2020differentiable, vishniakou2021differentiable}. 
Many biological samples show features close to the diffraction limit, such as axons or dendrites, which in this way can be exploited for aberration sensing. 
We demonstrated this under imaging conditions which are encountered with genetically encoded calcium indicators in fruit flies or zebrafish 
(Fig. \ref{fig:reults_1}, \ref{fig:results_2}).

Compared with other sensorless  methods that only use intensity characteristics or displacements of the image in one direction \cite{ji2010adaptive, hu2020universal}, the method developed here
relies on image modelling and allows to utilize a multidimensional feedback signal extracted from multiple pixels. However, in very dense and homogeneous samples, for example densely packed cell bodies that cover the entire field of view, the information that can be extracted in one measurement relies only on intensity, and in this situation feedback based methods are more suitable.

The phase output of DAO is not bound to any mode basis and can result in a phase surface with higher resolution than the spatial frequency of the probing modulations. We here nevertheless used Zernike modes to allow easier comparison with sensorless modal methods, requiring $2N+1$ \cite{debarre2009image} or $N+1$ measurements \cite{linhai2011wavefront, booth2006wave, booth2007wavefront}) for $N$ aberration modes.
We therefore also used the same number $N$ of probing modulations as Zernike modes present in the aberration.
DAO is additionally similar to another image based method which also requires $N$ measurements \cite{ji2010adaptive}, but in DAO the entire beam can be used for probing, thus increasing the available power and resolution for aberration sensing. 

The number of required probing modulations could potentially be further reduced using an optimal probing strategy by finding aberrations that would be suited best for gaining information about aberrations, and inferring the microscope-specific optimal probing mode basis as discussed in \cite{jesacher2011sensorless}. 
Probing of aberrations could be additionally improved by taking into account the three dimensional distribution of fluorescence typical observed in biological samples.  Fluorescent volumes generally lead to confounds for sensing aberrations with sensorless methods due to shifts in the focal plane or axial structure of the focus with changing probing modulations. This could be addressed by including a three dimensional sample distribution as an optimization parameter in DAO.

The simulations in Fig. \ref{fig:aberration_map} show the range of aberrations that one can expect to correct using DAO. For example, the distributions with fine and low intensity speckles will typically not be correctable using DAO due to the limited signal to noise ratio and dynamic range. 

The developed optimization approach can equally be used for light shaping (Fig.\ref{fig:result_hologram}). 
Since the model captures effects due to light polarization, it could similarly be applied in situations where such effects are of greater interest, for example for vectorial imaging or optical tweezers  \cite{torok2008high, foreman2011computational, chen2013holographic, liu2022super, kumar2022manipulation}. 

Overall, we have developed a differentiable optimization method for the Debye-Wolf integral and applied it for light shaping and adaptive optics. The provided software offers a unified framework  for efficiently modeling light propagation in high N.A. objectives and therefore can be used across microscopy applications.

\section*{Funding}
Max Planck Society, Max Planck Institute for Neurobiology of Behavior - caesar. 
 
\section*{Acknowledgments}
We thank Bahar Ghannad, Fabian Svara, Verena Juchems, Stefan Pauls, and Kevin Briggman for zebrafish samples. We thank Mitchell Sandoe for Scanimage support.
 
\section*{Disclosures}
None.

\section*{Data availability statement}
Data underlying the results presented in this paper are not publicly available at this time but may be obtained from the authors upon reasonable request.

\section*{Supplementary Materials}

1. \label{itm:vectorial_debye_model_foci} PSF of the model in response to the modulations and different polarizations. 2. \label{itm:model_matching_illustration} Illustration of model matching.

\bibliographystyle{unsrt}
\bibliography{submission}

\begin{thebibliography}{10}

\bibitem{bautista2016vector}
Godofredo Bautista and Martti Kauranen.
\newblock Vector-field nonlinear microscopy of nanostructures.
\newblock {\em ACS Photonics}, 3(8):1351--1370, 2016.

\bibitem{liu2022super}
Min Liu, Yunze Lei, Lan Yu, Xiang Fang, Ying Ma, Lixin Liu, Juanjuan Zheng, and
  Peng Gao.
\newblock Super-resolution optical microscopy using cylindrical vector beams.
\newblock {\em Nanophotonics}, 2022.

\bibitem{lutz2008holographic}
Christoph Lutz, Thomas~S Otis, Vincent DeSars, Serge Charpak, David~A
  DiGregorio, and Valentina Emiliani.
\newblock Holographic photolysis of caged neurotransmitters.
\newblock {\em Nature methods}, 5(9):821--827, 2008.

\bibitem{yang2011three}
Sunggu Yang, Eirini Papagiakoumou, Marc Guillon, Vincent De~Sars, Cha-Min Tang,
  and Valentina Emiliani.
\newblock Three-dimensional holographic photostimulation of the dendritic
  arbor.
\newblock {\em Journal of neural engineering}, 8(4):046002, 2011.

\bibitem{oron2012two}
Dan Oron, Eirini Papagiakoumou, Francesca Anselmi, and Valentina Emiliani.
\newblock Two-photon optogenetics.
\newblock {\em Progress in brain research}, 196:119--143, 2012.

\bibitem{chen2013holographic}
Hao Chen, Yunfeng Guo, Zhaozhong Chen, Jingjing Hao, Ji~Xu, Hui-Tian Wang, and
  Jianping Ding.
\newblock Holographic optical tweezers obtained by using the three-dimensional
  gerchberg--saxton algorithm.
\newblock {\em Journal of Optics}, 15(3):035401, 2013.

\bibitem{kumar2022manipulation}
Ram~Nandan Kumar, Subhasish~Dutta Gupta, Nirmalya Ghosh, and Ayan Banerjee.
\newblock Manipulation of mesoscopic particles using a structured beam in
  optical tweezers.
\newblock In {\em Mesophotonics: Physics and Systems at Mesoscale}, volume
  12152, pages 94--105. SPIE, 2022.

\bibitem{debarre2009image}
Delphine D{\'e}barre, Edward~J Botcherby, Tomoko Watanabe, Shankar Srinivas,
  Martin~J Booth, and Tony Wilson.
\newblock Image-based adaptive optics for two-photon microscopy.
\newblock {\em Optics letters}, 34(16):2495--2497, 2009.

\bibitem{ji2010adaptive}
Na~Ji, Daniel~E Milkie, and Eric Betzig.
\newblock Adaptive optics via pupil segmentation for high-resolution imaging in
  biological tissues.
\newblock {\em Nature methods}, 7(2):141--147, 2010.

\bibitem{hu2020efficient}
Yanlei Hu, Zhongyu Wang, Xuewen Wang, Shengyun Ji, Chenchu Zhang, Jiawen Li,
  Wulin Zhu, Dong Wu, and Jiaru Chu.
\newblock Efficient full-path optical calculation of scalar and vector
  diffraction using the bluestein method.
\newblock {\em Light: Science \& Applications}, 9(1):1--11, 2020.

\bibitem{wolf1959electromagnetic}
Emil Wolf.
\newblock Electromagnetic diffraction in optical systems-i. an integral
  representation of the image field.
\newblock {\em Proceedings of the Royal Society of London. Series A.
  Mathematical and Physical Sciences}, 253(1274):349--357, 1959.

\bibitem{zhang2021single}
Oumeng Zhang and Matthew~D Lew.
\newblock Single-molecule orientation localization microscopy i: fundamental
  limits.
\newblock {\em JOSA A}, 38(2):277--287, 2021.

\bibitem{foreman2008inversion}
Matthew~R Foreman, Sherif~S Sherif, Peter~RT Munro, and Peter T{\"o}r{\"o}k.
\newblock Inversion of the debye-wolf diffraction integral using an
  eigenfunction representation of the electric fields in the focal region.
\newblock {\em Optics Express}, 16(7):4901--4917, 2008.

\bibitem{torok2008high}
P~T{\"o}r{\"o}k, PRT Munro, and Em~E Kriezis.
\newblock High numerical aperture vectorial imaging in coherent optical
  microscopes.
\newblock {\em Optics express}, 16(2):507--523, 2008.

\bibitem{foreman2011computational}
Matthew~R Foreman and Peter T{\"o}r{\"o}k.
\newblock Computational methods in vectorial imaging.
\newblock {\em Journal of Modern Optics}, 58(5-6):339--364, 2011.

\bibitem{rabiner1969chirp}
L~Rabiner, R~W Schafer, and C~Rader.
\newblock The chirp z-transform algorithm.
\newblock {\em IEEE transactions on audio and electroacoustics}, 17(2):86--92,
  1969.

\bibitem{tensorflow2015-whitepaper}
Mart\'{i}n Abadi, Ashish Agarwal, Paul Barham, Eugene Brevdo, Zhifeng Chen,
  Craig Citro, Greg~S. Corrado, Andy Davis, Jeffrey Dean, Matthieu Devin,
  Sanjay Ghemawat, Ian Goodfellow, Andrew Harp, Geoffrey Irving, Michael Isard,
  Yangqing Jia, Rafal Jozefowicz, Lukasz Kaiser, Manjunath Kudlur, Josh
  Levenberg, Dandelion Man\'{e}, Rajat Monga, Sherry Moore, Derek Murray, Chris
  Olah, Mike Schuster, Jonathon Shlens, Benoit Steiner, Ilya Sutskever, Kunal
  Talwar, Paul Tucker, Vincent Vanhoucke, Vijay Vasudevan, Fernanda Vi\'{e}gas,
  Oriol Vinyals, Pete Warden, Martin Wattenberg, Martin Wicke, Yuan Yu, and
  Xiaoqiang Zheng.
\newblock {TensorFlow}: Large-scale machine learning on heterogeneous systems,
  2015.
\newblock Software available from tensorflow.org.

\bibitem{de2018end}
Filipe de~Avila Belbute-Peres, Kevin Smith, Kelsey Allen, Josh Tenenbaum, and
  J~Zico Kolter.
\newblock End-to-end differentiable physics for learning and control.
\newblock {\em Advances in neural information processing systems}, 31, 2018.

\bibitem{schoenholz2020jax}
Samuel Schoenholz and Ekin~Dogus Cubuk.
\newblock Jax md: a framework for differentiable physics.
\newblock {\em Advances in Neural Information Processing Systems},
  33:11428--11441, 2020.

\bibitem{thuerey2021pbdl}
Nils Thuerey, Philipp Holl, Maximilian Mueller, Patrick Schnell, Felix Trost,
  and Kiwon Um.
\newblock {\em Physics-based Deep Learning}.
\newblock WWW, 2021.

\bibitem{hanser2004phase}
Bridget~M Hanser, Mats~GL Gustafsson, DA~Agard, and John~W Sedat.
\newblock Phase-retrieved pupil functions in wide-field fluorescence
  microscopy.
\newblock {\em Journal of microscopy}, 216(1):32--48, 2004.

\bibitem{booth2006wave}
Martin~J Booth.
\newblock Wave front sensor-less adaptive optics: a model-based approach using
  sphere packings.
\newblock {\em Optics express}, 14(4):1339--1352, 2006.

\bibitem{linhai2011wavefront}
Huang Linhai and Changhui Rao.
\newblock Wavefront sensorless adaptive optics: a general model-based approach.
\newblock {\em Optics express}, 19(1):371--379, 2011.

\bibitem{thao2020phase}
Nguyen~Hieu Thao, Oleg Soloviev, and Michel Verhaegen.
\newblock Phase retrieval based on the vectorial model of point spread
  function.
\newblock {\em JOSA A}, 37(1):16--26, 2020.

\bibitem{vishniakou2020differentiable}
Ivan Vishniakou and Johannes~D Seelig.
\newblock Differentiable model-based adaptive optics with transmitted and
  reflected light.
\newblock {\em Optics Express}, 28(18):26436--26446, 2020.

\bibitem{vishniakou2021differentiable}
Ivan Vishniakou and Johannes~D Seelig.
\newblock Differentiable model-based adaptive optics for two-photon microscopy.
\newblock {\em Optics Express}, 29(14):21418--21427, 2021.

\bibitem{booth1998aberration}
Martin~J Booth, Mark~AA Neil, and Tony Wilson.
\newblock Aberration correction for confocal imaging in
  refractive-index-mismatched media.
\newblock {\em Journal of microscopy}, 192(2):90--98, 1998.

\bibitem{song2010model}
H~Song, R~Fraanje, G~Schitter, H~Kroese, G~Vdovin, and M~Verhaegen.
\newblock Model-based aberration correction in a closed-loop
  wavefront-sensor-less adaptive optics system.
\newblock {\em Optics express}, 18(23):24070--24084, 2010.

\bibitem{booth2007wavefront}
Martin~J Booth.
\newblock Wavefront sensorless adaptive optics for large aberrations.
\newblock {\em Optics letters}, 32(1):5--7, 2007.

\bibitem{sherman2002adaptive}
L~Sherman, Jing~Yong Ye, O~Albert, and Theodore~B Norris.
\newblock Adaptive correction of depth-induced aberrations in multiphoton
  scanning microscopy using a deformable mirror.
\newblock {\em Journal of microscopy}, 206(1):65--71, 2002.

\bibitem{marsh2003practical}
P~N Marsh, D~Burns, and J~M Girkin.
\newblock {Practical implementation of adaptive optics in multiphoton
  microscopy}.
\newblock {\em Optics express}, 11(10):1123--1130, 2003.

\bibitem{wright2005exploration}
Amanda~J Wright, David Burns, Brett~A Patterson, Simon~P Poland, Gareth~J
  Valentine, and John~M Girkin.
\newblock Exploration of the optimisation algorithms used in the implementation
  of adaptive optics in confocal and multiphoton microscopy.
\newblock {\em Microscopy Research and Technique}, 67(1):36--44, 2005.

\bibitem{vellekoop2015feedback}
Ivo~M Vellekoop.
\newblock Feedback-based wavefront shaping.
\newblock {\em Optics express}, 23(9):12189--12206, 2015.

\bibitem{emiliani2022optogenetics}
Valentina Emiliani, Emilia Entcheva, Rainer Hedrich, Peter Hegemann, Kai~R
  Konrad, Christian L{\"u}scher, Mathias Mahn, Zhuo-Hua Pan, Ruth~R Sims,
  Johannes Vierock, et~al.
\newblock Optogenetics for light control of biological systems.
\newblock {\em Nature Reviews Methods Primers}, 2(1):1--25, 2022.

\bibitem{Gerchberg72}
R.~W. Gerchberg and W.~O. Saxton.
\newblock {Practical algorithm for determination of phase from image and
  diffraction plane pictures}.
\newblock {\em {Optik}}, {35}({2}):{237--\&}, {1972}.

\bibitem{pozzi2018fast}
Paolo Pozzi, Laura Maddalena, Nicol{\`o} Ceffa, Oleg Soloviev, Gleb Vdovin,
  Elizabeth Carroll, and Michel Verhaegen.
\newblock Fast calculation of computer generated holograms for 3d
  photostimulation through compressive-sensing gerchberg--saxton algorithm.
\newblock {\em Methods and protocols}, 2(1):2, 2018.

\bibitem{shen2006fast}
Fabin Shen and Anbo Wang.
\newblock Fast-fourier-transform based numerical integration method for the
  rayleigh-sommerfeld diffraction formula.
\newblock {\em Applied optics}, 45(6):1102--1110, 2006.

\bibitem{pologruto2003scanimage}
Thomas~A Pologruto, Bernardo~L Sabatini, and Karel Svoboda.
\newblock Scanimage: flexible software for operating laser scanning
  microscopes.
\newblock {\em Biomedical engineering online}, 2(1):1--9, 2003.

\bibitem{boruah2009focal}
BR~Boruah and MAA Neil.
\newblock Focal field computation of an arbitrarily polarized beam using fast
  fourier transforms.
\newblock {\em Optics communications}, 282(24):4660--4667, 2009.

\bibitem{leutenegger2006fast}
Marcel Leutenegger, Ramachandra Rao, Rainer~A Leitgeb, and Theo Lasser.
\newblock Fast focus field calculations.
\newblock {\em Optics express}, 14(23):11277--11291, 2006.

\bibitem{goodman2005introduction}
Joseph~W Goodman.
\newblock {\em Introduction to Fourier optics, 3rd ed.}
\newblock Roberts and Company Publishers, 2005.

\bibitem{mendoza2014encoding}
Omel Mendoza-Yero, Gladys M{\'\i}nguez-Vega, and Jes{\'u}s Lancis.
\newblock Encoding complex fields by using a phase-only optical element.
\newblock {\em Optics letters}, 39(7):1740--1743, 2014.

\bibitem{gross2007handbook}
Herbert Gross, Z~Hannfried, Martin Peschka, Fritz Blechinger, et~al.
\newblock {\em Handbook of Optical Systems, Volume 3: Aberration Theory and
  Correction of Optical Systems}.
\newblock Wiley-Vch, 2007.

\bibitem{zhang2021fast}
Yan Zhang, M{\'a}rton R{\'o}zsa, Yajie Liang, Daniel Bushey, Ziqiang Wei,
  Jihong Zheng, Daniel Reep, Gerard~Joey Broussard, Arthur Tsang, Getahun
  Tsegaye, et~al.
\newblock Fast and sensitive gcamp calcium indicators for imaging neural
  populations.
\newblock {\em Biorxiv}, 2021.

\bibitem{jenett2012gal4}
Arnim Jenett, Gerald~M Rubin, Teri-TB Ngo, David Shepherd, Christine Murphy,
  Heather Dionne, Barret~D Pfeiffer, Amanda Cavallaro, Donald Hall, Jennifer
  Jeter, et~al.
\newblock A gal4-driver line resource for drosophila neurobiology.
\newblock {\em Cell reports}, 2(4):991--1001, 2012.

\bibitem{flores2022dynamics}
Andres Flores-Valle and Johannes~D Seelig.
\newblock Dynamics of a sleep homeostat observed in glia during behavior.
\newblock {\em bioRxiv}, 2022.

\bibitem{ahrens2012brain}
Misha~B Ahrens, Jennifer~M Li, Michael~B Orger, Drew~N Robson, Alexander~F
  Schier, Florian Engert, and Ruben Portugues.
\newblock Brain-wide neuronal dynamics during motor adaptation in zebrafish.
\newblock {\em Nature}, 485(7399):471--477, 2012.

\bibitem{hu2020universal}
Q~Hu, J~Wang, J~Antonello, M~Hailstone, M~Wincott, R~Turcotte, D~Gala, and
  MJ~Booth.
\newblock A universal framework for microscope sensorless adaptive optics:
  Generalized aberration representations.
\newblock {\em APL Photonics}, 5(10):100801, 2020.

\bibitem{tang2012superpenetration}
Jianyong Tang, Ronald~N Germain, and Meng Cui.
\newblock Superpenetration optical microscopy by iterative multiphoton adaptive
  compensation technique.
\newblock {\em Proceedings of the National Academy of Sciences},
  109(22):8434--8439, 2012.

\bibitem{papadopoulos2017scattering}
Ioannis~N Papadopoulos, Jean-S{\'e}bastien Jouhanneau, James~FA Poulet, and
  Benjamin Judkewitz.
\newblock Scattering compensation by focus scanning holographic aberration
  probing (f-sharp).
\newblock {\em Nature Photonics}, 11(2):116--123, 2017.

\bibitem{may2021fast}
Molly~A May, Nicolas Barr{\'e}, Kai~K Kummer, Michaela Kress, Monika
  Ritsch-Marte, and Alexander Jesacher.
\newblock Fast holographic scattering compensation for deep tissue biological
  imaging.
\newblock {\em Nature Communications}, 12(1):1--8, 2021.

\bibitem{jesacher2011sensorless}
Alexander Jesacher and Martin~J Booth.
\newblock Sensorless adaptive optics for microscopy.
\newblock In {\em MEMS Adaptive Optics V}, volume 7931, pages 115--123. SPIE,
  2011.

\end{thebibliography}

\end{document}